\documentclass[twocolumn,showpacs,preprintnumbers,amsmath,amssymb]{revtex4}
\usepackage{graphicx}
\usepackage{dcolumn}
\usepackage{bm}
\usepackage[latin1]{inputenc}

\begin{document}

\title{Fermionic bound states on a one-dimensional lattice}
\author{Jean-Pierre Nguenang$^{1,2}$}
\author{Sergej Flach$^{1}$}
\affiliation{1\ Max-Planck-Institut f\"ur Physik komplexer Systeme, N\"othnitzer Str. 38, 01187 Dresden, Germany}
\affiliation{2\ Fundamental Physics Laboratory: Group of Nonlinear physics and Complex
systems, Department of Physics, University of Douala, P.O. Box 24157, Douala, Cameroon }

\date{\today}

\begin{abstract}
We study bound states of two fermions with opposite spins in an extended Hubbard chain.
The particles interact when located both on a site or on adjacent sites.
We find three different types of bound states. 
Type U is predominantly formed of basis states with both fermions on the same site,
while two states of type V originate from both fermions occupying neighbouring sites.
Type U, and one of the states from type V, are symmetric
with respect to spin flips. The remaining one from type V is antisymmetric.
V-states are characterized by a diverging localization length below some critical wave number.
All bound states become compact for wave numbers at the edge of the Brilloin zone. 

\end{abstract}


\maketitle

{\sl Introduction.}
Advances in experimental techniques of manipulation of ultracold atoms in optical
lattices make it feasible to explore the physics of few-body interactions. Systems with few quantum particles
on lattices have new unexpected features as compared to the condensed matter case of
many-body interactions, where excitation energies are typically small compared to the Fermi energy.
In particular, a recent experiment explored the repulsive binding of bosonic atom pairs
in an optical lattice \cite{Winkler2006Nature441}, as predicted theoretically decades earlier \cite{aao69,EilbeckPhysicaD78}
(see also \cite{fg08} for a review).
In this work, we study binding properties of fermionic pairs with total spin zero.
We use the extended Hubbard model, which contains two interaction scales - the on site interaction $U$ and
the nearest neighbour intersite interaction $V$. The nonlocal interaction $V$ is added in condensed matter physics to emulate
remnants of the Coulomb interaction due to non-perfect screening of electronic charges.
For fermionic ultracold atoms or molecules with magnetic or electric 
dipole-dipole interactions, it can be tuned with respect to
the local interaction $U$ by modifying the trap geometry of a condensate, additional
external dc electric fields, combinations with fast rotating external fields, etc (for a review
and relevant references see \cite{mab08}).

The paper is organized as follows. We first describe the model
and introduce the basis we use to
write down the Hamiltonian matrix to be diagonalized. Then we derive the quantum
states of the lattice containing one and two fermions, with opposite spins in the latter case.  
We study the obtained bound states for two fermions. 
We obtain analytical expressions for the energy spectrum
of the bound states.
Some of the bound states are characterized by a critical momentum below which they dissolve
with the two-particle continuum.

{\sl Model and basis choice.}
We consider a one-dimensional lattice with
$f$ sites and periodic boundary conditions described by the extended Hubbard model with the following Hamiltonian:
\begin{equation}\label{eq:hamiltonian}
\hat{H} = \hat{H}_0 +  \hat{H}_U + \hat{H}_V,
\end{equation}
where
\begin{equation}
\hat{H}_0=-\sum_{j,\sigma} \hat{a}^+_{j,\sigma} (\hat{a}_{j-1,\sigma} + \hat{a}_{j+1,\sigma} )\;,
\end{equation}
\begin{equation}\label{eq:haminteraction1}
\hat{H}_U = -U\sum_{j} \hat{n}_{j,\uparrow} \hat{n}_{j,\downarrow}\;,\;\hat{n}_{j,\sigma}=
\hat{a}^+_{j,\sigma} \hat{a}_{j,\sigma}\;,
\end{equation}
\begin{equation}\label{eq:haminteraction2}
\hat{H}_V = -V\sum_{j} \hat{n}_j \hat{n}_{j+1}\;,\; \hat{n}_j = \hat{n}_{j,\uparrow}+ \hat{n}_{j,\downarrow}\;.  
\end{equation}
$\hat{H}_0$ describes the nearest-neighbor hopping of fermions
along the lattice. Here the symbols $\sigma = \uparrow,\downarrow$ stand for
a fermion with spin up or down.  $\hat{H}_U$
describes the onsite interaction between the particles, and 
$\hat{H}_V$  the intersite interaction of fermions located at adjacent sites.
$ \hat{a}^+_{j,\sigma}$ and $\hat{a}_{j,\sigma}$ are the fermionic creation and
annihilation operators satisfying the corresponding anticommutation relations: 
$[\hat{a}^+_{j,\sigma},\hat{a}_{l,\sigma'}]=\delta_{j,l}\delta_{\sigma,\sigma'}$,
$[\hat{a}^+_{j,\sigma},\hat{a}^+_{l,\sigma'}]=[\hat{a}_{j,\sigma},\hat{a}_{l,\sigma'}]=0$.
Note that throughout this work we consider $U$ and $V$ positive, which leads to bound states
located below the two-particle continuum. A change of the sign of $U,V$ will simply swap the energies.

The Hamiltonian (\ref{eq:hamiltonian}) commutes with the number operator
$\hat{N}=\sum_{j}\hat{n}_j$  whose
eigenvalues are $n=n_{\uparrow}+n_{\downarrow}$, i.e. the total number of fermions in
the lattice. We consider $n=2$,
with $n_{\uparrow}=1$ and $n_{\downarrow}=1$.
Therefore we construct a basis starting with the eigenstates of $\hat{N}$.
We use a number state basis $|\Phi_n\rangle=|n_1;n_2\cdots n_f\rangle$ 
\cite{EilbeckPhysicaD78}, where $n_i=n_{i,\uparrow}+ n_{i,\downarrow}$
represents the number of fermions  at the i-th site of the lattice. 
$|\Phi_n\rangle$ is an eigenstate of the number operator $\hat{N}$ with
eigenvalue $n=\sum_{j=1}^f n_j$. 

To observe the fermionic character of the considered states, any two-particle number state
is generated from the vacuum $|O\rangle$ by first creating a particle with spin down,
and then a particle with spin up: e.g. $\hat{a}^+_{2,\uparrow} \hat{a}^+_{1,\downarrow} |O\rangle$
creates a particle with spin down on site 1 and one with spin up on site 2, while 
$\hat{a}^+_{2,\uparrow} \hat{a}^+_{2,\downarrow} |O\rangle$ creates both particles
with spin down and up on site 2.

Due to periodic boundary conditions the Hamiltonian (\ref{eq:hamiltonian}) commutes also with the 
translation operator $\hat{T}$, which shifts all lattice indices by one.
It has eigenvalues $\tau=exp(ik)$, with Bloch wave number $k=\frac{2\pi\nu}{f}$ and $\nu=0,1,2,...,f-1$.

{\sl Single particle spectrum.}
For the case of having only one fermion (either spin up or spin down) in the lattice ($n=1$),
a number state has the form $|j\rangle = \hat{a}^+_{j,\sigma} |O\rangle$.
The interaction terms $\hat{H}_U$ and
$\hat{H}_V$  do not contribute.
For a given wave number $k$, the eigenstate to (\ref{eq:hamiltonian}) is therefore given by:
\begin{equation}
|\Psi_1\rangle=\frac{1}{\sqrt{f}} \sum_{s=1}^f\Big(\frac{\hat{T}}{\tau}\Big)^{s-1}|1\rangle\;.
\end{equation}
The corresponding eigenenergy 
\begin{equation}\label{eq:1fermionenergy}
\varepsilon_k=-2\cos(k).
\end{equation}

{\sl Two fermions with opposite spins.}
For two particles,  the number state method involves $N_{s}= f^2$  basis states,
which is the number of ways one can distribute two
fermions with opposite spins over the $f$ sites including possible double occupamcy of a site.
Below we consider only cases of odd $f$ for simplicity. Extension to even values of $f$ is straightforward.

We define basis states to a given value of the wave number $k$:
\begin{eqnarray}
|\Phi_1\rangle = \frac{1}{\sqrt{f}} \sum_{s=1}^f\Big(\frac{\hat{T}}{\tau}\Big)^{s-1}
\hat{a}^+_{1,\uparrow} \hat{a}^+_{1,\downarrow} |O\rangle\;, \\
|\Phi_{j,+}\rangle = \frac{1}{\sqrt{f}} \sum_{s=1}^f\Big(\frac{\hat{T}}{\tau}\Big)^{s-1}
\hat{a}^+_{j,\uparrow} \hat{a}^+_{1,\downarrow} |O\rangle\;,\\
|\Phi_{j,-}\rangle = \frac{1}{\sqrt{f}} \sum_{s=1}^f\Big(\frac{\hat{T}}{\tau}\Big)^{s-1}
\hat{a}^+_{1,\uparrow} \hat{a}^+_{j,\downarrow} |O\rangle\;, \\
\; 1 < j \leq (f+1)/2\;.
\end{eqnarray}
Note, that the sign $\pm$ discriminates between states, where the distance from the spin up particle
to the spin down particle is smaller (larger) than vice versa. Note that distances are measured
by scanning the (periodic) chain in the direction of increasing lattice site number, passing a given particle
(say with spin up) and then counting the distance to the spin down particle. In the limit of an infinite system
the two signs discriminate between states where the spin up particle is to the left or right of the spin down one.

Therefore, a complete wavefunction is given by 
 \begin{equation}\label{eq:basis1}
|\Psi_2\rangle= c_1|\Phi_1\rangle+\sum_{j=2}^{\frac{f+1}{2}}c_{j+}|\Phi_{j,+}\rangle
              +\sum_{j=2}^{\frac{f+1}{2}}c_{j-}|\Phi_{j,-}\rangle\;.
\end{equation}
Any vector in our given Hilbert space is then spanned by the numbers $|c_1,c_{2+},c_{2-},c_{3+},...\rangle$.

Next we calculate the matrix elements of the Hamiltonian
(\ref{eq:hamiltonian}) in the framework of the  basis (\ref{eq:basis1}).
We arrive at a $f \times f$ matrix with elements
$H(i,j)$ ($i,j=1,\ldots,d$):
\begin{eqnarray}\label{eq:matel2}
H(i,j) = -\left( \begin{array}{cccccccc}
U & q &q & 0 & \ldots & ~& ~& \\
q^* & V&0& q & 0 & \ldots & & \\
q^* &0&V&0&q & 0 & \ldots  & \\
0 &q^* & 0 & 0 & 0 & q  & 0 &  \\
0 &0 &\ddots&\ddots &\ddots &\ddots &\ddots & 0 \\
  &\ldots&0&q^*&0&0&0&q\\
  & & \ldots & 0 & q^* & 0 & 0 & p \\
  & & &\ldots & 0 &q^*& p & 0
\end{array} \right)\;.
\end{eqnarray}
Here $q=1+\tau$ and $p=\tau^{-(f+1)/2} +
\tau^{-(f-1)/2}$. 

\begin{figure}
\includegraphics[width=0.95\columnwidth]{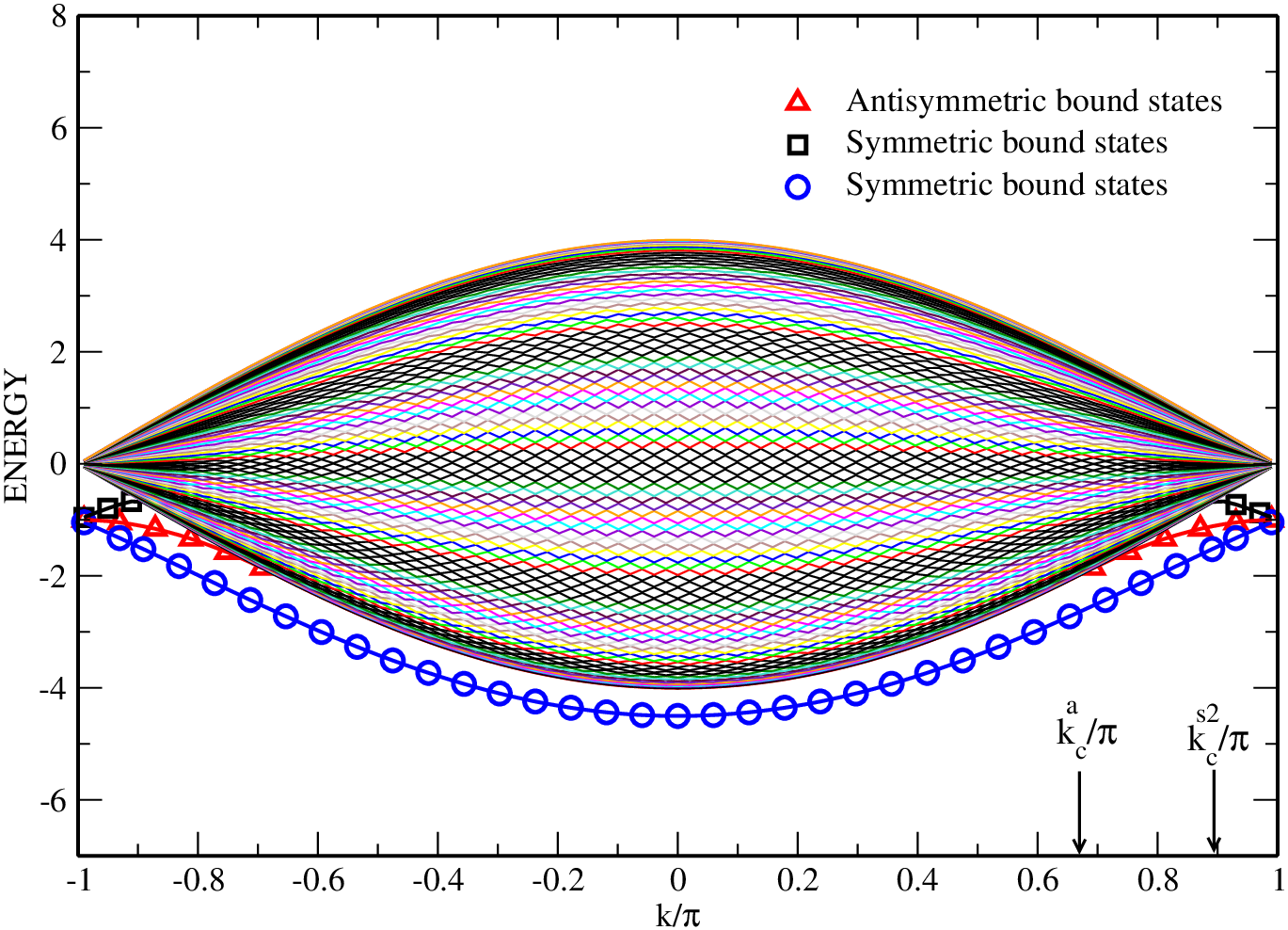}
\caption{\label{spectrum}Energy spectrum of the two fermion states.
The eigenvalues are plotted as a function of the wave
number $k$. Here $U=1\;,\;V=1,\;f=101$. 
Symbols are
from analytical derivation, lines are the result of numerical diagonalization.
The arrows indicate the location of the critical wave numbers (see text).
}
\end{figure} 

In Fig. \ref{spectrum} we show the energy spectrum of the
Hamiltonian matrix (\ref{eq:matel2}) obtained by numerical diagonalization for
the interaction parameters $U=2$ and  $V=2$ and $f=101$. 
At $U=0$ and $V=0$, the spectrum is given by the two fermion continuum, whose
eigenstates are
characterized by the two fermions independently moving along the
lattice. In this case the eigenenergies are the sum of the two single-particle energies:
\begin{equation}\label{eq:energyunperturbed}
E_{k_1,k_2}^0=-2[\cos(k_1)+\cos(k_2)],
\end{equation}
with $k_{1,2}={\pi}\nu_{1,2}/(f+1)$ , 
$\nu_{1,2}= 1,\ldots,f$. 
The Bloch wave number $k = k_1+k_2 \mod 2\pi$. Therefore, if $k=\pm \pi$,
the continuum degenerates into points.
The continuum is bounded by the hull curves $h_{\pm}(k)=\pm 4 \cos \frac{k}{2}$.
The same two-particle continuum is still observed in 
Figure.\ref{spectrum} for nonzero interaction.
However, in addition to the continuum, we observe one, two or three bound states
dropping out of the continuum, which depends on the wave number.
For any nonzero $U$ and $V$, all three bound states drop out of the 
continuum at $k=\pm \pi$.
One of them stays bounded for all values of $k$. The two other ones merge with the
continuum at some critical value of $|k|$ upon approaching $k=0$ as observed in
Fig.\ref{spectrum}. Note that for $k=\pm \pi$ and $U=V$, all three bound states
are degenerate.

\begin{figure}
\includegraphics[width=0.95\columnwidth]{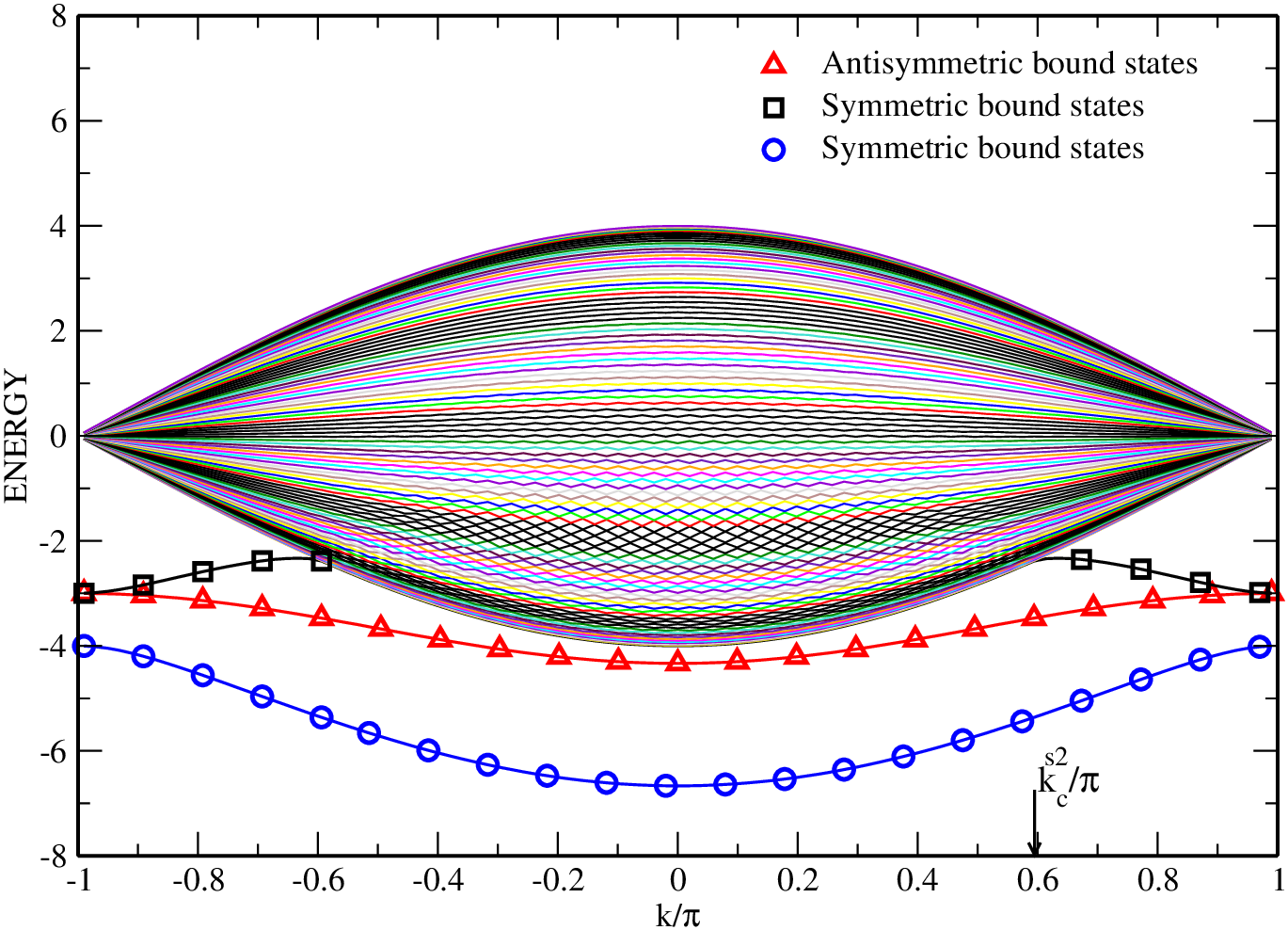}
\caption{\label{pbcspectrum}Energy spectrum of the two fermion states.  
The eigenvalues are plotted as a function of the wave
number k.  Here $U=4\;,\;V=3\;,\;f=101$. The symbols are
from analytical derivation, lines are the results of numerical diagonalization.
The arrow indicates the location of the critical wave number (see text).
}
\end{figure}

Upon increasing $U$ and $V$, we observe
that a second bound state band separates from the continuum for all $k$ (Fig.\ref{pbcspectrum}). At the same time,
when $U \neq V$,
the degeneracy at $k=\pm \pi$ is reduced to two.
 
\begin{figure}
\includegraphics[width=0.95\columnwidth]{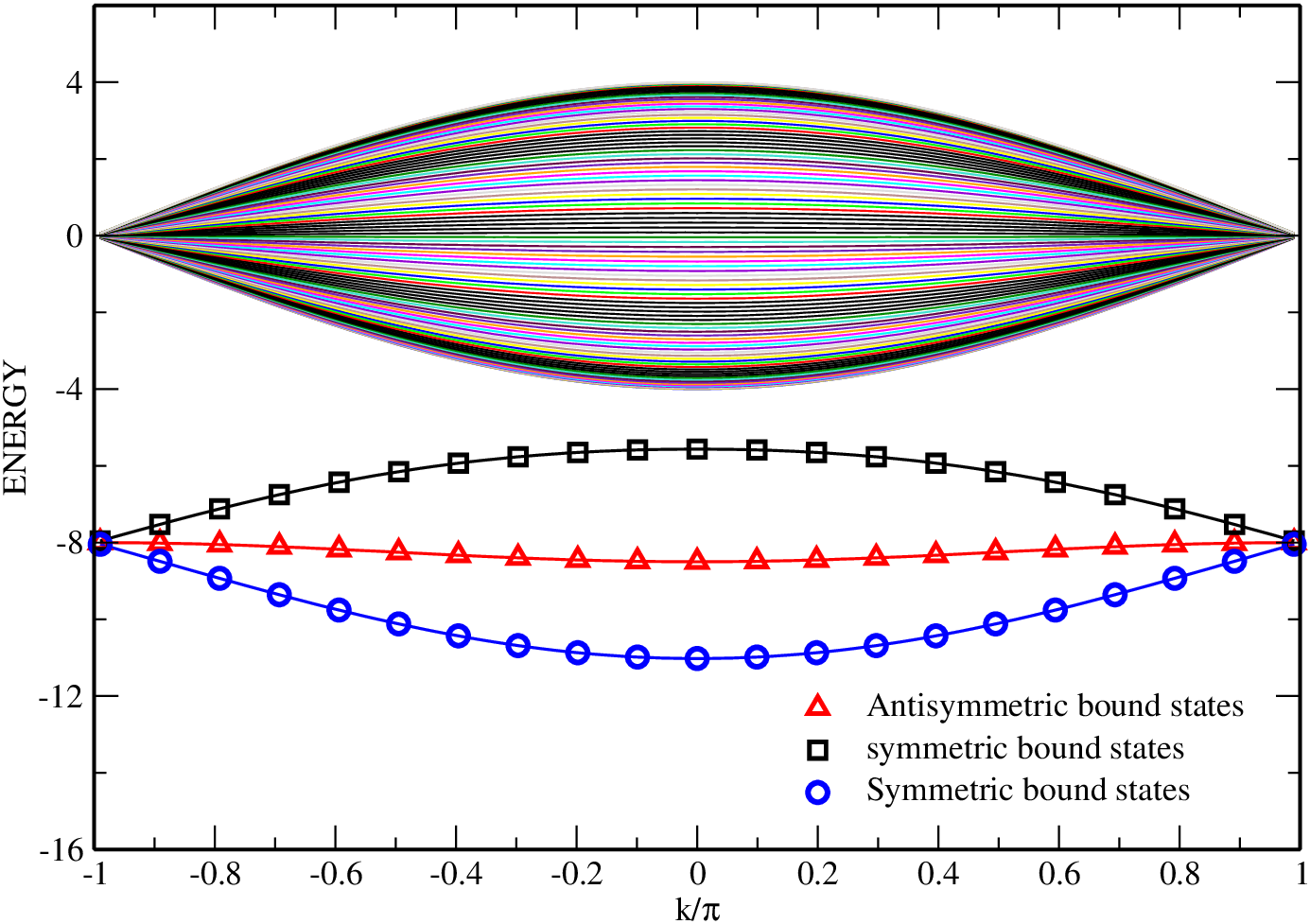}
\caption{\label{symspect} Energy spectrum for the two fermion states.
The eigenvalues are plotted as a function of the wave
number k.  Here $U=8\;,\;V=8\;,\;f=101$. The symbols are
from analytical derivation, lines are the results of numerical diagonalization.
}
\end{figure}

Finally, for even larger values of $U$ and $V$, all three bound state bands completely
separate from the continuum (Fig.\ref{symspect}).

{\sl Symmetric and antisymmetric state representation.}
In order to obtain analytical estimates on the properties of the observed bound states,
we use the fact that the Hamiltonian for a two fermion state is invariant under flipping
the spins of both particles. 
We define symmetric basis states
\begin{equation}\label{eq:basisym}
|\Phi_{j,s}\rangle=\frac{1}{\sqrt{2}}(|\Phi_{j,+}\rangle + |\Phi_{j,-}\rangle)
\end{equation}
and antisymmetric states
\begin{equation}\label{eq:basisas}
|\Phi_{j,a}\rangle=\frac{1}{\sqrt{2}}(|\Phi_{j,+}\rangle - |\Phi_{j,-}\rangle)\;.
\end{equation}
Note that $|\Phi_1\rangle$ is a symmetric state as well.
Then the matrix (\ref{eq:matel2}) can be decomposed into irreducible symmetric and antisymmetric
ones:
\begin{eqnarray}\label{eq:matsym}
H^s(i,j) = -\left( \begin{array}{cccccc}
U & q\sqrt{2} & & & & \\
q^*\sqrt{2} &V & q & & & \\
      & q^* & 0 & q & & \\
  &  & \ddots & \ddots &\ddots & \\
  &  & & q^* & 0 & q \\
  &  & &  & q^* & p
\end{array} \right)\;,
\end{eqnarray}
\begin{eqnarray}\label{eq:matasym}
H^a(i,j) = -\left( \begin{array}{cccccc}
V & q& & & & \\
q^* & 0 & q & & & \\
      & q^* & 0 & q & & \\
  &  & \ddots & \ddots &\ddots & \\
  &  & & q^* & 0 & q \\
  &  & &  & q^* &- p
\end{array} \right)\;.
\end{eqnarray}
The rank of the matrix $H^s$ is $(f+1)/2$, while the rank of $H^a$ is $(f-1)/2$. 

{\sl Antisymmetric bound states.}
The antisymmetric states exclude double occupation. Therefore the spectrum is identical
with the one of two spinless fermions \cite{EilbeckPhysicaD78}.
Following the derivations in \cite{EilbeckPhysicaD78}
we find that the antisymmetric bound state, if it exists, has an energy
\begin{equation}
\label{eanti}
E^a_2(k)=-(V+\frac{4}{V}\cos^2(\frac{k}{2}))\;.
\end{equation}
This result is valid as long as the the bound state energy stays outside of the 
continuum. The critical value of $k$ at which validity is lost, is obtained
by requesting $|E^a_2(k)|=|h_{\pm}(k)|$. 
It follows
$ V=2\cos(\frac{k}{2})$. 
Therefore the antisymmetric bound state merges with the continuum at a critical wave number
\begin{equation}
\label{kac}
 k^a_{c}=2\arccos(\frac{V}{2})\;,
\end{equation} 
setting a critical length scale $\lambda^a_{c}=\frac{2\pi}{ k^a_{c}}$.
For $V=1$ it follows $k^a_{c}/\pi\approx 0.667$ (see Fig.\ref{spectrum}).

The equation (\ref{eanti}) is in excellent agreement with the numerical data in 
Figs. \ref{spectrum},\ref{pbcspectrum},\ref{symspect} (cf. open triangles). We also note, that the
antisymmetric bound state is located between the two symmetric bound states, which we
discuss next. 

{\sl Symmetric bound states.}
A bound state can be searched for by assuming an unnormalized eigenvector to (\ref{eq:matsym}) of the form
$|c,1,\mu,\mu^2,\mu^3,...\rangle$ with $|\mu| \equiv \rho \leq 1$.
We obtain
\begin{eqnarray}
E c = U c + \sqrt{2} q^*\;, \nonumber \\
E = \sqrt{2} q c + V + q^* \mu \;, \\
E = \frac{1}{\mu} + q^* \mu \;. \nonumber
\end{eqnarray}
It follows that $\mu = \rho \exp(ik/2)$ and
\begin{equation}{\label{eq:ensym}}
E^s_2(k) = -2({\rho+\frac{1}{\rho})\cos{k/2}}\;.
\end{equation}
The parameter $\rho$ satisfies a cubic equation
\begin{equation}\label{eq:cubiceq}
 a{\rho}^3+b{\rho}^2+c{\rho}+d=0
\end{equation}
with the real coefficients a, b. c and d given by
$a=2Vcos(\frac{k}{2})$, 
$b= 4cos^2(\frac{k}{2})-UV$,
$c=2(U+V)cos(\frac{k}{2})$,
$d=-4cos^2({\frac{k}{2}})$.
An analytic solution to (\ref{eq:cubiceq}) can be obtained, but is cumbersome
to be presented here.
We plot the results in Figs. \ref{spectrum},\ref{pbcspectrum},\ref{symspect} (cf. open circles
and squares). We obtain excellent agreement.

At the Brilloin zone edge $k=\pm \pi$ the cubic equation (\ref{eq:cubiceq})
is reduced to a quadratic one, and can be solved to obtain finally $\rho \rightarrow 0$ and
\begin{equation}{\label{eq:enrhosmal}}
E^{s1}_{2}(k\to\pm\pi)=-U\;,\;E^{s2}_{2}(k\to\pm\pi) =-V \;.
\end{equation}
In particular we find for $k=\pm \pi$ that $E^{s2}_{2} = E^a_2$. In addition, if $U=V$,
all three bound states degenerate at the zone edge.

If $V=0$, the cubic equation (\ref{eq:cubiceq}) is reduced to a quadratic one in the whole
range of $k$ and yields \cite{EilbeckPhysicaD78}
\begin{equation}
E^{s1}_{2}(k) = -\sqrt{U^2 + 16 \cos^2 ( k/2)}\;.
\end{equation}

Next we determine the critical value of $k$ for which the bound state with energy $E^{s2}_{2}$
is joining the continuum. Since at this point $\rho=1$, we solve (\ref{eq:cubiceq})
with respect to $k_c$ and find
\begin{equation}
\label{ks2c}
k^{s2}_{c}=2\arccos\Big( \frac{UV}{2(U+2V)}\Big)
\end{equation}
setting another critical length scale $\lambda^s_{c}=\frac{2\pi}{ k^s_{c}}$.
E.g. for $U=V=1$ $k^{s2}_{c}/\pi \approx 0.89$, in excellent agreement with Fig.\ref{spectrum}.
For $U=4$ and $V=3$ we find $k^{s2}_{c}/\pi \approx 0.59$ confirming numerical results in Fig.\ref{pbcspectrum}.

{\sl Conclusions.}
Two fermionic particles with opposite spin allow for three different types of bound states
on a one-dimensional lattice with onsite $U$ and nearest neighbour $V$ interaction. 
Two of them are symmetric with respect to spin flips, and one
is antisymmetric. The antisymmetric bound state is characterized by a critical
wave number separates wave numbers with bound states from wave numbers without.
It follows from (\ref{kac}) that this happens for $V < 2$. For larger values of $V$ the whole
wave number space becomes available for antisymmetric bound states,
similar to one of the symmetric bound states for any nonzero $U$.
The second symmetric bound state also observes a critical wave number.
It follows from (\ref{ks2c}) that this happens for $U < 4V/(V-2)$,
while the whole wave number space becomes available otherwise.
It could be a challenging task to observe these different phases with one, two, or
three bound states experimentally, by tuning $U$, $V$, and $k$.

For higher lattice dimensions more nearest neighbours have to be taken into account,
similar to an increase of the interaction range.
In these cases, we expect consequently more bound states to appear.
\\
\\
Acknowledgements
\\
We thank M. Haque, D, Krimer, A. Ponno and Ch. Skokos for useful discussions.
J.-P. Nguenang acknowledges the warm hospitality of the Max
Planck Institute for the Physics of Complex Systems in Dresden.

\end{document}